\documentclass[12pt]{iopart}

\usepackage{iopams}
\usepackage{bm}
\usepackage{cite}
\usepackage{mathrsfs}

\newcommand{\ppi}{\bm{\mathnormal{\Pi}}}
\newcommand{\mno}{\mathnormal{\Omega}}
\newcommand{\mng}{\mathnormal{\Gamma}}

\begin{document}

\title[Instabilities of electromagnetic pulses]{Modulational and filamentational 
instabilities of two electromagnetic pulses in a radiation background} 

\author{Mattias Marklund, Padma K Shukla, Gert Brodin, and Lennart 
Stenflo}

\address{Department of Physics, Ume{\aa} University, SE--901 87
Ume{\aa}, Sweden}

\date{\today}

\begin{abstract}
  The nonlinear interaction, due to quantum
  electrodynamical (QED) effects, between two electromagnetic pulses and
  a radiation gas is investigated. It is found that the governing
  equations admit both modulational and  
  filamentational instabilities. The instability growth rates are
  derived, and the results are discussed. 
\end{abstract}



\section*{}

Within quantum electrodynamics (QED), there are new and interesting
phenomena with no classical counterparts. A prominent example 
is the Casimir effect. Similarly,
photon--photon scattering (see, e.g., 
\cite{Heisenberg-Euler,Weisskopf,Schwinger,Greiner-Muller-Rafaelski})
mediated by virtual electron--positron
pairs, does no occur within classical electrodynamics, in
which electromagnetic waves in vacuum are indifferent to each other. The
collisions of photons with photons have attracted much interest
over the years, both from an experimental and an astrophysical point
of view (see 
\cite{Bialynicka-Birula,Adler,Harding,Ding-Kaplan1,Latorre-Pascual-Tarrach,%
  Dicus-Kao-Repko,Ding-Kaplan2,Soljacic-Segev,Brodin-etal,%
  Brodin-marklund-Stenflo,Boillat,Heyl-Hernquist,%
  DeLorenci-Klippert-Novello,Thoma,Marklund-Brodin-Stenflo,Yu} 
and references therein). The effect of 
photon--photon scattering could be of fundamental importance in 
high-intensity laser pulses, in ultra-strong cavity fields, in the
surrounding of neutron 
stars and magnetars, and in the early Universe. 

In this paper we shall investigate the interaction between two
electromagnetic pulses via a background radiation gas. The gas acts as
a mediator of the interaction, and the pulses exchange their energy
through this background. Moreover, in contrast to the case with no
radiation 
background, even parallel propagation of the pulses results in 
non-zero coupling. It is furthermore shown that the resulting system
of equations lead to both modulational and filamentational
instabilities. Here we present the growth rates for these
processes. Our results are finally discussed in the context of
astrophysical and laboratory applications. 

We express the nonlinear self-interaction of the photons in terms
of the Heisenberg--Euler Lagrangian as found in Ref.\
\cite{Heisenberg-Euler} 
\begin{equation}\label{eq:lagrangian}
  L = \varepsilon_0F +
  \kappa\varepsilon_0^2\left[4F^2 + 7G^2 \right],
\end{equation}
where $F = (E^2 - c^2B^2)/2$ and $G = c\bm{E}\cdot\bm{B}$.
Here $\kappa \equiv 2\alpha^2\hbar^3/45m_e^4c^5 \approx 1.63\times
10^{-30}\, \mathrm{m}\mathrm{s}^{2}/\mathrm{kg}$, $\alpha$ is
the fine-structure constant, $\hbar$ is the Planck constant divided by 
$2\pi$, $m_e$ is the electron mass, and $c$ is the speed of light in vacuum. 
The Lagrangian (\ref{eq:lagrangian}) is valid when there is no
electron--positron  
pair creation, and when the field strength is smaller than the
critical field, i.e.  
\begin{equation}
  \omega \ll m_ec^2/\hbar , \, |\bm{E}| \ll
  E_{\mathrm{crit}} \equiv 
  m_ec^2/e\lambda_c   
  \label{eq:constraint} 
\end{equation}
respectively. Here $e$ is the elementary charge, $\lambda_c$ is the
Compton wavelength, and $E_{\mathrm{crit}} \simeq
10^{18}\,\mathrm{V}/\mathrm{m}$.   

Following Ref. \cite{Bialynicka-Birula}, we write the dispersion relation
for a low energy photon in a background electromagnetic field
($\bm{E}$, $\bm{B}$) as (see also Ref.~\cite{Boillat} and references therein)
\begin{equation}\label{eq:dispersionrelation}
  \omega(\bm{k}, \bm{E}, \bm{B}) = c|\bm{k}|\left( 1 -
  \case{1}{2}\lambda|\bm{Q}|^2 \right) ,
\end{equation}
where
\begin{eqnarray}
  |\bm{Q}|^2 = \varepsilon_0\left[ E^2 + c^2B^2 - 
   (\hat{\bm{k}}\cdot\bm{E})^2 -
   c^2(\hat{\bm{k}}\cdot\bm{B})^2 -
   2c\hat{\bm{k}}\cdot(\bm{E}\times\bm{B})\right] ,
\label{eq:Q2}
\end{eqnarray}
and $\lambda = \lambda_{\pm}$, where $\lambda_+ = 14\kappa$ and
$\lambda_- = 8\kappa$ for the two different polarisation states of the
photon. Furthermore, $\hat{\bm{k}} \equiv \bm{k}/k$.
The approximation $\lambda|\bm{Q}|^2 \ll 1$ has been used.

We will below study (a) plane electromagnetic pulses
propagating in a background consisting of a radiation gas in
equilibrium, and (b) a radiation gas affected by an electromagnetic
(EM) pulse propagating through the gas.

The electromagnetic field in Eq.\ (\ref{eq:Q2}) is here a
superposition of the fields 
due to the two coherent electromagnetic pulses, with electric fields
$\bm{E}_{1,2}(\bm{r},t)\exp[i(\bm{k}_{1,2}\cdot\bm{r} - \omega_{1,2} t)] +$
c.c., and the incoherent field due to the
photon gas. Here c.c.\ denotes complex conjugate. The amplitudes of the pulses
are supposed to be slowly varying as compared to their rapidly
oscillating phases. 
In the absence of the pulse, the photon gas is assumed to be isotropic. The
validity of the Lagrangian requires that the electromagnetic fields are
smaller than the critical field. Thus we can use a perturbative
approach where the 
photon gas is only weakly perturbed, i.e., to lowest order in a wave amplitude
expansion, the photon gas is still isotropic. 

Now we need to establish the connection between the dispersion
relation (\ref{eq:dispersionrelation}) 
(originally derived for static or slowly varying background fields), and the
contribution from the dynamical radiation fields of our problem. Here
it is crucial 
that the QED nonlinearities do not involve the derivatives of the
electromagnetic fields. In particular, we may write the general wave equation
resulting from the Heisenberg--Euler Lagrangian as
\begin{equation}
\left(\frac{1}{c^2}\frac{\partial^2}{\partial t^2}  - \nabla^2\right)
E_{i} = \sum_{j,k,l = 1}^3 \left( f_{ijkl}E^{j}E^{k}E^{l} +
g_{ijkl}B^{j}B^{k}E^{l} + h_{ijkl}B^{j}E^{k}E^{l} \right) . 
\end{equation}
The specific forms of the tensors $f,g$ and $h$ can be found by comparison
with the wave equations in for example Refs.\
\cite{Soljacic-Segev,Brodin-marklund-Stenflo}. We note that the
nonlinear terms are cubic, i.e.\ the lowest order scattering processes
involve four waves. 
The coefficient $|\bm{Q}|^2$ in (\ref{eq:Q2}) results from
carrying out all possible summations, where one of the fields in
the right hand side is taken as the wave field, and the others as the
constant fields, and finally taking the scalar product with the polarization
vector. However, for combinatorial reasons, the coefficient of interaction becomes the same
for waves influenced by background fields that are dynamical waves. This can be seen by noting that the 
combination surviving the wave mixing is obtained by taking one field as the original wave, the other as a wave field at a
different frequency and the third as its complex conjugate
\footnote{One 
notes that an overall factor two enters, because of the
contributions from the terms where the wave field changes place with the
complex conjugate term. However, this is compensated for by using the time
averaged values of the fields rather than the peak amplitudes.}. Other combinations surviving the wave mixing 
are ruled out by the conservation relations.

Next we substitute the particular fields 
into (\ref{eq:Q2}) in order to evaluate their effects on the
dispersion relation (\ref{eq:dispersionrelation}). 
Here we note that the rapidly oscillating contributions to
$|\bm{Q}|^2$ should not be included, i.e., terms proportional to
$\bm{E}_1\cdot \bm{E}_2 \exp
[i((\bm{k}_1 + \bm{k}_2)\cdot\bm{r} - (\omega_1 + \omega_2)t)]$,
$\bm{E}_1^2 \exp [2i(\bm{k}_1\cdot \bm{r} - \omega_1 t)]$ etc. can be
neglected, since 
in Eq.\ (\ref{eq:dispersionrelation}) such terms will be averaged
out. Thus the only terms kept in the expression for $|\bm{Q}|^2$ are
proportional either to the photon gas density or to the slowly varying
intensity in one of the pulses. We note that our calculation of the
effect of the 
pulses on the photon gas requires the intensities to vary on a long scale as
compared to the wavelengths of the gas photons. 

For case (a), the relations 
$ (\hat{\bm{k}}_i\cdot\bm{E})^2 = E^2/3$,
  $(\hat{\bm{k}}_i\cdot\bm{B})^2 = B^2/3$, 
and $\bm{E}\times\bm{B} = 0$
hold, where $\bm{k}_i$, $i = 1,2$, stands for the pulse wavevectors.
Hence, from (\ref{eq:Q2}) we obtain
\begin{eqnarray}
  |\bm{Q}_1|^2 = \case{4}{3}\mathscr{E}_g + \alpha_1\mathscr{E}_2 ,
  \label{eq:Q2gas1} \\ 
  |\bm{Q}_2|^2 = \case{4}{3}\mathscr{E}_g + \alpha_2\mathscr{E}_1 ,
  \label{eq:Q2gas2} 
\end{eqnarray}
where $\mathscr{E}_g = \varepsilon_0(E^2 + c^2B^2)/2$ and
$\mathscr{E}_i = \varepsilon_0|E_i|^2$ is the energy
density of the radiation gas and pulse $i$, respectively. Using 
$\hat{\bm{e}}_i = \bm{E}_i/|E_i|$, we have defined
\begin{eqnarray}
  \alpha_{1,2} = 2 -
   2\hat{\bm{k}}_{1}\cdot\hat{\bm{k}}_{2} -
   (\hat{\bm{k}}_{1,2}\cdot\hat{\bm{e}}_{2,1})^2 -
   [(\hat{\bm{k}}_{1,2}\times\hat{\bm{k}}_{2,1})\cdot\hat{\bm{e}}_{2,1})]^2
   .
\end{eqnarray}

In case (b), the directions $\hat{\bm{k}}_g$ of the photons in the 
gas are approximately random. The last statement holds true as long as
the pulse energy densities are small, thus inducing only weak
deviations from the equilibrium values of the radiation gas
quantities. The error in this approximation is therefore only of first
order in the pulse energy densities, giving a second order correction
in the final equations. Then (\ref{eq:Q2}) yields
\begin{equation}\label{eq:Q2pulse}
  |\bm{Q}_g|^2 =
   \case{4}{3}\left( \mathscr{E}_1 + \mathscr{E}_2 +
   \beta\sqrt{\mathscr{E}_1\mathscr{E}_2}  \right) ,
\end{equation}
where we have introduced the parameter
\begin{equation}
  \beta = \hat{\bm{e}}_1\cdot\hat{\bm{e}}_2 +
   (\hat{\bm{k}}_1\cdot\hat{\bm{k}}_2)\hat{\bm{e}}_1\cdot\hat{\bm{e}}_2
   -
   (\hat{\bm{k}}_1\cdot\hat{\bm{e}}_2)(\hat{\bm{k}}_2\cdot\hat{\bm{e}}_1) . 
\end{equation}

With the relations (\ref{eq:Q2gas1})--(\ref{eq:Q2gas2}), we use standard methods for
slowly varying envelopes \cite{Hasegawa} to derive the
dynamical equations for two pulses on a photon gas background, to
obtain 
\begin{eqnarray}
\fl i\left( \frac{\partial}{\partial t} +
  c\hat{\bm{k}}_{01}\cdot\nabla
  \right)\bm{E}_1 +
  \frac{c}{2k_{01}}\left[ \nabla^2 -
  (\hat{\bm{k}}_{01}\cdot\bm{\nabla})^2\right]\bm{E}_1 +
  \lambda ck_{01}\left( \frac{2}{3}\mathscr{E}_g +
  \frac{1}{2}\alpha_1\mathscr{E}_2 \right)\bm{E}_1 = 0 , 
  \label{eq:nlse1} \\
\fl i\left( \frac{\partial}{\partial t} +
  c\hat{\bm{k}}_{02}\cdot\nabla
  \right)\bm{E}_2 +
  \frac{c}{2k_{02}}\left[ \nabla^2 -
  (\hat{\bm{k}}_{02}\cdot\bm{\nabla})^2\right]\bm{E}_2 +
  \lambda ck_{02}\left( \frac{2}{3}\mathscr{E}_g +
  \frac{1}{2}\alpha_2\mathscr{E}_1 \right)\bm{E}_2 = 0 ,
  \label{eq:nlse2}
\end{eqnarray}
where the index $0$ refers to the non-interacting background
state. We note that if the two pulses are co-propagating, their mutual
interactions disappear, while for counter-propagating pulses we have
$\alpha_i = 4$.

Following Ref.\ \cite{Marklund-Brodin-Stenflo} and using
(\ref{eq:Q2pulse}), we then derive the
equations for the radiation gas on a background of two
plane wave pulses (assuming that the rate of change in the direction
of the pulses is slow)
\begin{eqnarray}
  &&\fl \frac{\partial\mathscr{E}_g}{\partial t} +
  c^2\bm{\nabla}\cdot\bm{\ppi} = \frac{4}{3}\lambda\left[
  \left( 1 +
  \frac{\beta}{2}\sqrt{\frac{\mathscr{E}_2}{\mathscr{E}_1}}
  \right)\left( -\frac{\mathscr{E}_g}{2}\frac{\partial}{\partial t} +
  c^2{\ppi}\cdot\bm{\nabla}\right)\mathscr{E}_1 \right. \nonumber \\
  &&\fl\left. \quad + \left( 1 +
  \frac{\beta}{2}\sqrt{\frac{\mathscr{E}_1}{\mathscr{E}_2}}
  \right)\left( -\frac{\mathscr{E}_g}{2}\frac{\partial}{\partial t} +
  c^2{\ppi}\cdot\bm{\nabla}\right)\mathscr{E}_2
 + c^2\left( \mathscr{E}_1 + \mathscr{E}_2 +
  \beta\sqrt{\mathscr{E}_1\mathscr{E}_2 } \right)\bm{\nabla}\cdot{\ppi}
  \right],
  \label{eq:fluid1}
\end{eqnarray}
and
\begin{eqnarray}
  \frac{\partial{\ppi}}{\partial t} + 
  \frac{1}{3}\bm{\nabla}\mathscr{E}_g =
  \frac{2}{3}\lambda\mathscr{E}_g \left[ \left( 1 +
  \frac{\beta}{2}\sqrt{\frac{\mathscr{E}_2}{\mathscr{E}_1}}
  \right)\bm{\nabla}\mathscr{E}_1 
  + \left( 1 +
  \frac{\beta}{2}\sqrt{\frac{\mathscr{E}_1}{\mathscr{E}_2}}
  \right)\bm{\nabla}\mathscr{E}_2 
  \right] ,
  \label{eq:fluid2}
\end{eqnarray}
where ${\ppi}$ is the momentum density of the photon gas. The coupled
equations (\ref{eq:nlse1})--(\ref{eq:nlse2}) and 
(\ref{eq:fluid1})--(\ref{eq:fluid2}) describe the momentum
exchange due to wave--wave scattering of the high frequency
pulses and the low frequency gas photons.   
We note that
Eqs.\ (\ref{eq:fluid1})--(\ref{eq:fluid2}) have been derived by taking moments of the Vlasov
equation for the radiation gas photons. We have chosen to start from
an effective Lagrangian for the photon--photon interaction, but one
could in principle also start from the scattering cross section of the
elastic photon--photon collisions, forming a collisional integral of
the Boltzmann equation, and from there derive the fluid
equations. However, for the purposes of this paper, the Vlasov
approach is sufficient.  

Introducing $\mathscr{E}_g = \mathscr{E}_0 + \delta\mathscr{E}$, where
$\mathscr{E}_0$ ($\gg \delta\mathscr{E}$) is the unperturbed radiation
gas energy density in the absence of the pulses, and assuming that the
unperturbed momentum density of the gas vanishes, we obtain from Eqs.\
(\ref{eq:fluid1})--(\ref{eq:fluid2}) 
\begin{eqnarray}
  && 
  \frac{\partial^2\delta\mathscr{E}}{\partial t^2} -
  \frac{c^2}{3}\nabla^2\delta\mathscr{E} =
  -\frac{2}{3}\lambda\mathscr{E}_0 \Bigg\{ 
  \left( 1 +
  \frac{\beta}{2}\sqrt{\frac{\mathscr{E}_2}{\mathscr{E}_1}}
  \right)\left( \frac{\partial^2}{\partial t^2} + c^2\nabla^2
  \right)\mathscr{E}_1  \nonumber \\ &&
  + \left( 1 +
  \frac{\beta}{2}\sqrt{\frac{\mathscr{E}_1}{\mathscr{E}_2}}
  \right)\left( \frac{\partial^2}{\partial t^2} + c^2\nabla^2
  \right)\mathscr{E}_2 \nonumber \\ && 
%
  - \frac{\beta}{4}\sqrt{\frac{\mathscr{E}_2}{\mathscr{E}_1^3}} 
  \left[ \left( \frac{\partial\mathscr{E}_1}{\partial t}\right)^2 +
  c^2|\nabla\mathscr{E}_1|^2  \right]
  - \frac{\beta}{4}\sqrt{\frac{\mathscr{E}_1}{\mathscr{E}_2^3}} 
  \left[ \left( \frac{\partial\mathscr{E}_2}{\partial t}\right)^2 +
  c^2|\nabla\mathscr{E}_2|^2  \right] \nonumber \\ &&
%
  + \frac{\beta}{2\sqrt{\mathscr{E}_1\mathscr{E}_2}}
  \left[ \frac{\partial\mathscr{E}_1}{\partial
  t}\frac{\partial\mathscr{E}_2}{\partial t} +
  c^2(\bm{\nabla}\mathscr{E}_1)\cdot(\bm{\nabla}\mathscr{E}_2) \right] 
  \Bigg\} 
\label{eq:wave}
\end{eqnarray}
Equation (\ref{eq:wave}) together with Eqs.\ 
(\ref{eq:nlse1})--(\ref{eq:nlse2}) form a
Karpman-like system of equations
\cite{Karpman1,Zakharov,Karpman2,Karpman3}.

In order to analyse modulational instabilities, we proceed along the
lines of Ref.\ \cite{Shukla} and let  
$
  \delta\mathscr{E} = \widetilde{\delta\mathscr{E}}\exp[{i(\bm{K}\cdot{r}
  - {\mno} t)}]$ and $\mathscr{E}_j =
  \widetilde{\mathscr{E}}_j\exp[{i(\bm{K}\cdot{r} 
  - {\mno} t)}]$ where $j = 1, 2$. 
From Eq.\ (\ref{eq:wave}) we then obtain
\begin{equation}
  \widetilde{\delta\mathscr{E}} =
  \frac{2}{3}\lambda\mathscr{E}_0\Delta\left( \widetilde{\mathscr{E}}_1 +
  \widetilde{\mathscr{E}}_2 +
  \beta\sqrt{\widetilde{\mathscr{E}}_1\widetilde{\mathscr{E}}_2}
  \right) ,
\label{eq:pert}
\end{equation}
where $\Delta = ({\mno}^2 + c^2K^2)/(-{\mno}^2 + c^2K^2/3)$.

In the absence of the radiation gas, the interaction between
co-propagating pulses vanishes. This is not the case when the
radiation gas is present, as will now be shown. 
For parallel propagation, we have 
$\hat{\bm{k}}_{01} = \hat{\bm{k}}_{02}= \hat{\bm{k}}_{0}$, and
$\nabla^2 - (\hat{\bm{k}}_{0}\cdot\bm{\nabla})^2 =
\nabla_{\perp}^2$. Furthermore, parallel pulses imply $\alpha_1 =
\alpha_2 = 0$, i.e., the Eqs.\ (\ref{eq:nlse1})--(\ref{eq:nlse2}) take the form
\begin{eqnarray}
  && i\left( \frac{\partial}{\partial t} +
  c\hat{\bm{k}}_{0}\cdot\nabla
  \right)\bm{E}_j 
  +
  \frac{c}{2k_{0j}}\nabla_{\perp}^2\bm{E}_j \nonumber \\ 
  &&\qquad  + 
   \frac{2}{3}\lambda ck_{0j}\mathscr{E}_0
  \left[ 1 +
  \frac{2}{3}\lambda\varepsilon_0\Delta\left(|E_1|^2 +
  |E_2|^2   +
  \beta|E_1||E_2| \right) \right]\bm{E}_j = 0  ,
\label{eq:new-nlse}  
\end{eqnarray} 
where we have used Eq.\ (\ref{eq:pert}). The
modulational instability of Eqs.\ (\ref{eq:new-nlse}) will now be
analysed, using the ansatz 
$
  \bm{E}_j = (\bm{E}_{j0} + \widetilde{\bm{E}}_j)e^{i\varphi_jt}
$, 
where $|\widetilde{\bm{E}}_j| \ll |\bm{E}_{j0}|$, and where
$\bm{E}_{j0}$ is a 
real constant vector. To lowest order, Eq.\ (\ref{eq:new-nlse})
gives the frequency shift 
$
  \varphi_j = \frac{2}{3}\lambda ck_{0j}\mathscr{E}_0\left[ 1 +
  \frac{2}{3}\lambda\varepsilon_0\Delta\left( E_{10}^2 +
  E_{20}^2 + \beta E_{10}E_{20} \right) \right] ,
$ 
and to next order we obtain 
\begin{eqnarray}
 && i\left( \frac{\partial}{\partial t} +
  c\hat{\bm{k}}_{0}\cdot\nabla
  \right)\widetilde{\bm{E}}_j +
  \frac{c}{2k_{0j}}\nabla_{\perp}^2\widetilde{\bm{E}}_j 
 \nonumber \\
 &&\quad  +
  \varepsilon_0 W_j\left[ 
     (\widetilde{\bm{E}}_1 + \widetilde{\bm{E}}_1^*)\cdot\bm{E}_{10} +
     (\widetilde{\bm{E}}_2 + \widetilde{\bm{E}}_2^*)\cdot\bm{E}_{20} 
   \right]\bm{E}_{j0} = 0 ,
\label{eq:lse}
\end{eqnarray}
where the star denotes the complex conjugate, and $W_j =
(4/9)\lambda^2ck_{0j}\mathscr{E}_0\Delta$.  

To further simplify the analysis, we assume that
$\widetilde{\bm{E}}_j\cdot\bm{E}_{j0} = \widetilde{E}_j{E}_{j0}$,
and that the polarisation of the perturbation vectors remain fixed to
lowest order, and set $k_{01} = k_{02} = k_0$. We let 
$
  \widetilde{E}_j = (X_j + i Y_j)\exp[i(\bm{K}\cdot\bm{r} - {\mno} t)] 
$, 
and Fourier analyse Eq.\ (\ref{eq:lse}) to obtain 
\begin{eqnarray}
\fl  ({\mno} - cK_{\|})^4 - \frac{cK_{\perp}^2}{k_0}\left[
  \frac{cK_{\perp}^2}{2k_0} - W\mathscr{E}_p
  \right]({\mno} -cK_{\|})^2 
  + \frac{c^3K_{\perp}^6}{4k_0^3}\left[
  \frac{cK_{\perp}^2}{4k_0} - W\mathscr{E}_p
  \right]  = 0 ,
\end{eqnarray}
where $W = W_1 = W_2$, $\mathscr{E}_p
= \varepsilon_0(E_{10}^2 + E_{20}^2)$, $K_{\|} =
\hat{\bm{k}}_0\cdot\bm{K}$ and $K_{\perp}^2 = K^2 - K_{\|}^2$. Thus,
for ${\mno} = cK_{\|} + i{\mng}$ and ${\mng} \ll cK_{\|}$, we find
that  
\begin{equation}
 {\mng} \simeq \frac{1}{2}{cK_{\perp}}\sqrt{
 \frac{16}{3}\eta_0\eta_p\frac{K_{\perp}^2
 + 2K_{\|}^2}{K_{\perp}^2 - 2K_{\|}^2} - \frac{K_{\perp}^2}{k_0^2}} ,
\end{equation}
where the dimensionless parameters $\eta_0 = \lambda\mathscr{E}_0$
and $\eta_p = \lambda\mathscr{E}_p$ give the coupling strength for
the radiation gas and the pulse, respectively. Thus, we see that the
radiation gas acts as a mediator for the energy transfer between the
two pulses, something that is not possible with the pulses alone. 
The positivity of ${\mng}^2$ requires that 
\begin{equation}
  K_{\|}^2 > \frac{K_{\perp}^2}{2}\frac{(3K_{\perp}^2 -
  16k_0^2\eta_0\eta_p)}{(3K_{\perp}^2 +
  16k_0^2\eta_0\eta_p)} \geq 0 .
\end{equation}
We furthermore notice that the growth rate diverges when $K_{\|}$
approaches the ``critical'' value $\pm K_{\perp}/\sqrt{2}$, 
at which the approximation ${\mng} \ll cK_{\|}$ breaks down. 
On the other hand, even if the parameters $\eta_0$ and $\eta_p$ are small, 
one can nevertheless obtain a large growth rate, provided that $K_{\|}$ 
lies close to $\pm K_{\perp}/\sqrt{2}$. The phase speed at the critical 
wavenumber is $c/\sqrt{3}$, i.e. the speed of sound in a thermal radiation 
gas.

A stationary filamentation instability can likewise be obtained from
Eq.\ (\ref{eq:new-nlse}), analogously to the modulational
instability. If $K_{\|} = i\chi$, we obtain
\begin{equation}
  \chi = \frac{1}{2}K_{\perp}\sqrt{\frac{16}{3}\eta_0\eta_p -
  \frac{K_{\perp}^2}{k_0^2}} .
\end{equation}
Notice that, contrary to the modulational instability, the
filamentational instability growth rate does not have a singular
behaviour at certain parameter values.

The existence of a modulational and/or filamentational instability
within a physical system indicates that the nonlinear evolution,
either in time or in space, is nontrivial. When the
time-dependence is significant, the modulational instability implies
that any small perturbation will grow. We therefore expect the system
to evolve towards a  
strongly inhomogeneous state. Naturally higher order 
nonlinearities are likely to saturate such a collapse scenario. On
longer time-scales  
thermo-dynamical effects, outside the range of validity of our
equations, will become important.  
This is likely to eventually destroy the inhomogeneous structures.
Even if the system is stationary, the 
filamentational instability will single out a spatial direction along
which inhomogeneities will grow. Here, a full
nonlinear analysis requires numerical methods. 
We note that, in addition to the modulational and/or filamentational
instabilities considered here, four-wave scattering involving only 
high frequency photons is possible. However, for dimensional reasons
we expect the growth rates associated with such processes to be smaller
than the ones found here, provided that $\eta_0  >  \eta_p$. 

The situations in which photon--photon scattering can become important
are indeed in the extreme range of what Nature may offer. Although the
evolution of high power lasers has been rapid over the
last decades, we are still not near the limits where the direct effect
of lasers can give us important information of photon--photon
scatterings. Rather, this has to be sought by more indirect means, such
as the coherent interactions of cavity modes presented in Ref.\ 
\cite{Brodin-marklund-Stenflo}, or in astrophysical environments, such
as close to magnetars (see, e.g., Ref.\ \cite{magnetar}), where field
strengths close to the 
Schwinger limit may exist, and effects such as photon splitting or
magnetic lensing, as found in Ref.\
\cite{Bialynicka-Birula,Adler,Harding,Heyl-Hernquist}, may take place. 
Moreover, the possibility of using recent high
precision measurements of the cosmic microwave background, such as
the Wilkinson Microwave Anisotropy Probe (see Ref.\ \cite{wmap}), for
the indirect detection of photon--photon 
scatterings has also been suggested in Ref.\ \cite{Marklund-Brodin-Stenflo}.

In this paper we have suggested a theory for pairs of
electromagnetic pulses propagating on a radiation background. It has
been shown that the dynamics of the photon pulses is governed by a set of
coupled nonlinear Schr\"odinger equations, while the response of the
background can be deduced from an acoustic wave equation driven by the
pressure of  the 
two photon pulses. The full system resembles a Karpman set of equations. 
Moreover, we have investigated the interaction between two parallel
propagating  
electromagnetic pulses via a background radiation gas. The gas acts as
a mediator for the interaction, and the pulses exchange their energy
through this background. In contrast to the case of no radiation
background, when parallel propagation implies zero coupling, the
radiation gas in our case gives rise to a system
of equations possessing both modulational and filamentational
instabilities. We have presented the temporal and spatial amplification rates 
for both instabilities, and have discussed the implications of these
instabilities for the evolution of the system, both in laboratory
settings and in astrophysical environments.

\end{document}